%% file: template.tex
\documentclass[cameraready]{Interspeech}

\usepackage{cite}
\usepackage{amsmath,amssymb,amsfonts}
\usepackage{algorithmic}
\usepackage{graphicx}
\usepackage{textcomp}
\usepackage{xcolor}

\usepackage{hyperref}
\hypersetup{
    colorlinks=true,
    citecolor=blue,
}
\usepackage{colortbl}

\usepackage{cite}
\usepackage{url}
\usepackage{amsfonts}
\usepackage{multirow}
\usepackage{makecell}
\usepackage{booktabs} 
\usepackage{subcaption}

\title{CodecMOS-Accent: A MOS Benchmark of Resynthesized and TTS Speech from Neural Codecs Across English Accents}

\author[affiliation={1}, correspondingauthor]{Wen-Chin}{Huang}
\author[affiliation={2}]{Nicholas}{Sanders}
\author[affiliation={3}]{Erica}{Cooper}

\address{
    $^1$ Nagoya University, Japan \\
    $^2$ University of Edinburgh, United Kingdom \\
    $^3$ National Institute Of Information And Communications Technology, Japan
}

\email{wen.chinhuang@g.sp.m.is.nagoya-u.ac.jp}


\usepackage{comment}

\begin{document}

\maketitle

\begin{abstract}

We present the CodecMOS-Accent dataset, a mean opinion score (MOS) benchmark designed to evaluate neural audio codec (NAC) models and the large language model (LLM)-based text-to-speech (TTS) models trained upon them, especially across non-standard speech like accented speech. The dataset comprises 4,000 codec resynthesis and TTS samples from 24 systems, featuring 32 speakers spanning ten accents. A large-scale subjective test was conducted to collect 19,600 annotations from 25 listeners across three dimensions: naturalness, speaker similarity, and accent similarity. This dataset does not only represent an up-to-date study of recent speech synthesis system performance but reveals insights including a tight relationship between speaker and accent similarity, the predictive power of objective metrics, and a perceptual bias when listeners share the same accent with the speaker. This dataset is expected to foster research on more human-centric evaluation for NAC and accented TTS.
\end{abstract}

\section{Introduction}

Neural audio codecs (NACs) have become increasingly integral in modern speech processing systems \cite{recent-advances-in-discrete-speech-tokens, arora2025on, cui-etal-2025-recent, discrete-audio-tokens-more-than-a-survey}. Significant efforts have been directed toward establishing NACs as a more effective representation than alternatives such as continuous self-supervised learning (SSL)-based features \cite{speech-ssl-review}. These advancements range from simply tuning hyperparameters, such as codebook size and frame rate, to the application of sophisticated techniques like distillation, particularly as NACs can be strategically designed to isolate specific acoustic traits such as timbre or phonemes. Perhaps the most prominent advantage of NACs is their discrete nature, which facilitates the seamless modeling of speech within large language model (LLM) frameworks.

The growing interest in NACs has prompted researchers to establish benchmarking initiatives such as DASB \cite{DASB} and Codec-SUPERB \cite{codec-superb}. However, these benchmarks and individual NAC studies typically prioritize reconstruction quality and objective metrics for discriminative downstream tasks, often omitting subjective evaluations in the context of text-to-speech (TTS). Furthermore, most existing benchmarks utilize standard speech, leaving a gap in understanding how these codecs perform across diverse speech types, such as accented or emotional speech.

While naturalness is a standard dimension to evaluate in TTS, we are specifically interested in evaluating an emergent capability of LLM-based TTS: in-context learning (ICL). This refers to the ability to replicate various attributes of a reference speech sample, commonly known as the \textit{prompt}. This process, frequently termed ``voice cloning" in the literature \cite{neural-voice-cloning}, has been predominantly evaluated through the reproduction of \textit{speaker identity}. However, pioneering LLM-based TTS works like VALL-E \cite{vall-e} demonstrated that ICL can even replicate the acoustic recording environment. Consequently, researchers started to investigate whether these models can clone other attributes, including emotion, expressiveness, and accent \cite{cosyvoice3}. Despite this interest, such capabilities have yet to be systematically evaluated across a wide range of models. As a first step, this paper investigates how NACs and the LLM-based TTS models trained upon them generalize across diverse English accents.

Despite recent growing interests in synthesizing speech with the desired accent \cite{ac-tts-limited-data, accentbox, scalable-controllable-accented-tts}, the research community is still exploring the best way to evaluate accent similarity. While a recent study investigated whether perceptual accent similarity is reflected by commonly used objective metrics including classification results and cosine similarity of embeddings extracted from accent identification and speaker verification models \cite{pairwiase-evaluation-accent-sim}, their experiment remained limited in scale in terms of the number of accents, synthesis systems and listeners involved.

The CodecMOS-Accent dataset presented in this work was curated to meet the above-mentioned goals. This corpus comprises 4,000 samples from 24 contemporary codec resynthesis and TTS systems, featuring 32 speakers across ten distinct accents. We conducted a large-scale mean opinion score (MOS) test via a crowdsourcing company, recruiting 25 listeners from four self-reported accent regions to provide 19,600 annotations across three dimensions: naturalness, speaker similarity, and accent similarity. This dataset represents a comprehensive, up-to-date study of recent speech synthesis system performance. Our analysis sheds light on several critical research questions, reconfirming a strong correlation between subjective speaker and accent similarity, as well as the robust predictive power of objective metrics in this context. Furthermore, we identify a perceptual bias that occurs when listeners share the same accent as the speaker in the test sample. The dataset will be made public in the near future to further foster research on more perceptually guided speech quality assessment (SQA) methods for NAC and accented TTS research.

\section{The CodecMOS-Accent dataset}

\subsection{Material}

We chose VCTK \cite{vctk} to be our source material. Although other databases such as CommonVoice and AccentDB have been built for accented speech processing research, we found VCTK remains the best choice in terms of quality, quantity and diversity. We first downsampled all samples to 16kHz, and we trimmed the starting and trailing silences using open-sourced forced alignment labels\footnote{\url{https://github.com/kan-bayashi/VCTKCorpusFullContextLabel}}. We aimed to maintain a somewhat balanced distribution of accents, so using the accent labels in the VCTK dataset, we chose 32 speakers spanning 10 accents, with 20 female and 12 male speakers. We then chose 5 samples from each speaker, resulting in 160 ground truth samples in total. For the voice cloning task where a reference sample is needed, we randomly chose another different sample from the same speaker. We only chose samples from 3 to 7 seconds.

\subsection{Tasks and models}
\label{ssec:tasks-models}

The CodecMOS-Accent dataset contains 9 systems for the resynthesis task and 15 systems for the voice cloning task. Below we describe the task and models. Note that in this dataset, we only chose open-sourced models and excluded commercial, black-box models.

\subsubsection{Resynthesis}

In the resynthesis task, the NAC model processes the input speech by encoding it into a compressed representation and subsequently decoding it back into waveform. We selected the following models: Encodec \cite{encodec}, DAC \cite{dac}, SpeechTokenizer \cite{speechtokenizer}, FACodec \cite{naturalspeech3}, Mimi \cite{moshi}, SNAC \cite{snac}, WavTokenizer \cite{wavtokenizer}, NanoCodec \cite{nanocodec}, and NeuCodec \cite{neucodec}. Most modern NAC architectures adopt the residual vector quantization (RVQ) technique introduced in \cite{soundstream}, which enables flexible bitrate adjustment by utilizing only the initial layers of the codebook. To ensure a wide spectrum of audio quality within the dataset, we incorporated low-bitrate configurations for certain models.

\subsubsection{Voice cloning}

Voice cloning could be viewed as a specialized application of TTS. Given an input text and a reference audio sample, the objective is to generate speech that accurately articulates the target text while preserving both the speaker identity and accent present in the reference.
We selected the following models: VALL-E-X \cite{vall-e, valle-x}, TorToiSe \cite{tortoise}, XTTS \cite{xtts}, FireRedTTS \cite{fireredtts}, MaskGCT \cite{maskgct}, OpenAudio s1 mini \cite{openaudio}, VevoTTS \cite{vevo}, CosyVoice 2 \cite{cosyvoice2}, Llasa-1B \cite{llasa}, MetaVoice \cite{metavoice}, Orpheus-TTS \cite{orpheus-tts}, VoiceStar \cite{voicestar}, IndexTTS2 \cite{indextts2}, Chatterbox \cite{chatterboxtts2025}, and NeuTTS Air \cite{neuttsair}.

\subsection{Listening test design}

The listening test was conducted through Intergroup, a crowdsourcing company. We use the term \textit{annotation} to denote the tuple $\langle \text{test sample},\text{reference sample},\text{scores}\rangle$ from a specific listener. In each annotation there are three scores, described as follows:
\begin{itemize}
    \item \textbf{S-NAT}: listeners were asked to rate the naturalness of the test sample by considering pronunciation, prosody, noise, etc.
    \item \textbf{S-SPK-SIM}: listeners were asked to evaluate whether the test sample and the reference sample are spoken by the same speaker.
    \item \textbf{S-ACC-SIM}: listeners were asked to evaluate whether the test sample and the reference sample have the same accent, regardless of whether they sound like the same person.
\end{itemize}
All three scores are on a five-point scale. Finally, the listeners are also asked to write comments about the samples optionally.

The evaluated samples are from the 9 resynthesis and 15 TTS systems described in Section~\ref{ssec:tasks-models}, and the ground truth samples were also evaluated. Together there were 4,000 samples. We recruited 25 listeners who each rated 784 samples, resulting in a total of 19,600 anotations (4.9 annotations per sample.) No personal information was collected except for age, headphone information and self-reported accent. In total there were 19 US annotators, 2 Canadian annotators, 3 English annotators and 1 Scottish annotator\footnote{We followed VCTK and treat English and Scottish as different accents.}. Upon inspecting the comments, we rejected 55 samples (and the corresponding 275 annotations) reported to be of either only silence or very low quality such that fair evaluation was difficult to conduct.

\input{tables/results}
\input{tables/corr}
\input{tables/bias}
\input{tables/listener-group-corr-one-column}

\section{Analysis}

In this section we present the analysis of the collected CodecMOS-Accent dataset. In addition to the subjective scores, we also conducted objective evaluation using the following four commonly used metrics:

\begin{itemize}
    \item \textbf{O-WER}: word error rate, calculated using Whisper \cite{whisper}\footnote{\url{https://huggingface.co/openai/whisper-large-v3}}.
    \item \textbf{O-SPK-SIM}: cosine similarity of the speaker embeddings of the test sample and the reference sample, using an open-sourced ECAPA-TDNN model \cite{ecapa-tdnn}\footnote{\url{https://huggingface.co/speechbrain/spkrec-ecapa-voxceleb}}.
    \item \textbf{O-ACC-SIM}: cosine similarity of the accent embeddings of the test sample and the reference sample, using an open-sourced ECAPA-TDNN model trained on CommonAccent \cite{commonaccent}\footnote{\url{https://huggingface.co/Jzuluaga/accent-id-commonaccent_ecapa}}.
    \item \textbf{O-UTMOS}: predicted subjective speech quality, using UTMOS\cite{utmos}, the winning system of the VoiceMOS Challenge 2022 \cite{voicemos2022}. 
\end{itemize}

\subsection{System-level score analysis}

Our analysis starts by aggregating the scores on the system-level, and Table~\ref{tab:results} shows the results. We first observed that the ground truth ranked 9th in S-NAT. This is not surprising, as VCTK is known to contain acoustic artifacts and is frequently outperformed by modern TTS systems trained on high-fidelity datasets \cite{naturalspeech2} or those employing noise disentanglement techniques \cite{factorized-tacotron}. Conversely, the ground truth retained the highest rank in both S-SPK-SIM S-ACC-SIM, indicating that human evaluators can reliably assess speaker identity and accent independently of overall audio quality.

We also observed that compared to S-NAT with the lowest value being 1.36, both S-SPK-SIM and S-ACC-SIM have a higher score range, with lower bounds 2.84 and 3.13.
This discrepancy suggests that even when acoustic quality degrades, the evaluated models successfully preserve perceptible speaker and accent traits. This implies a learning hierarchy where models capture global attributes (speaker identity and accent) first, then learns to generate high-fidelity, artifact-free speech. A compelling demonstration of this phenomenon occurs in the resynthesized samples using the first two layers of the SpeechTokenizer. The persistence of speaker and accent traits in these low-layer samples directly contradicts the prevailing assumption that the initial layers of NAC models exclusively encode low-level linguistic features, such as pronunciation, while lacking high-level acoustic properties \cite{vall-e, speechtokenizer}.

Finally, we observed that the subjective scores of the resynthesis samples are upper-bounded by the ground truth. This shows the essence of the resynthesis task itself. On the other hand, many TTS systems outperformed the ground truth in terms of S-NAT. This shows that in the voice cloning task, some models only draws the essential information (speaker and accent) from the reference sample, without encoding the inherent environmental noise.

\subsection{Can exisiting objective metrics predict the subjective scores?}

We are interested in how predictive existing objective metrics are against subjective scores. So, we calculated the Pearson correlation coefficients between the subjective scores and the objective metrics. Table~\ref{tab:correlations} shows the results. First, we observed that S-SPK-SIM and S-ACC-SIM are highly correlated, even in the utterance-level (0.75).
This indicates that for the evaluated models, the capacity to accurately capture speaker identity is intrinsically linked to the synthesis of their corresponding accent.
On the system-level, S-NAT also shows moderate correlation with S-SPK-SIM (0.66) and S-ACC-SIM (0.59).
This suggests a trend where a model's proficiency in generating high-fidelity speech scales with its ability to synthesize specific speaker and accent characteristics.

Moving to the objective metrics, we first found that on the system level, O-SPK-SIM highly correlates with S-SPK-SIM (0.86), showing the high predictive power of O-SPK-SIM, similar to what was shown in early studies \cite{obj_vcc2020}. What was surprising was that O-SPK-SIM also correlated well with S-ACC-SIM (0.90), even better than O-ACC-SIM (0.81).
This shows that the ability to distinguish speakers automatically implies the ability to distinguish accents. We also found that O-WER had a much lower correlation with all subjective scores. While we acknowledge that O-WER remains a reliable measure of intelligibility, this finding strongly advocates against relying on O-WER as a solitary proxy for overall speech quality.

Perhaps the most surprising finding was the strong correlation coefficient of 0.96 between O-UTMOS and S-NAT. Developed in 2022, UTMOS was trained on the BVCC dataset \cite{bvcc}, which comprises synthesized speech from systems developed no later than 2020. Given that all the TTS systems evaluated in our study were introduced post-2020, the predictive power of UTMOS on these modern architectures is highly unexpected. However, this phenomenon highlights a critical caveat: algorithmic novelty does not inherently guarantee superior absolute synthesis quality. A significant portion of the BVCC dataset consists of high-quality, single-speaker TTS samples -- a task fundamentally less complex than the resynthesis and voice cloning tasks considered in this dataset. Consequently, it is highly plausible that contemporary NAC-based models, despite their advanced architectures, are currently undergoing the same development bottlenecks that single-speaker TTS systems faced a decade ago.

\subsection{Is accent a bias in subjective speech evaluation?}

In this part we are interested in whether the accent of the listener causes bias when evaluating accented speech. First, we compared the mean subjective scores from listeners who share the accent of the test sample against those from listeners with a different accent, and calculated $p$-values using Welch's t-test. Results for both ground truth samples alone and the full set of evaluated samples are presented in Table~\ref{tab:bias}.
For ground truth samples, we observed a statistically significant ``same-accent bias'' ($p < 0.05$) in both S-SPK-SIM and S-ACC-SIM, with listeners consistently assigning higher scores to speakers sharing their own accent. This suggests that listeners may be more generous when evaluating familiar accents. However, this bias was not present for S-NAT ($p > 0.05$), implying that naturalness judgments of real human speech remain relatively robust regardless of the listener's native accent.

Expanding our analysis to all samples, the same-accent bias persisted for S-SPK-SIM and S-ACC-SIM, but surprisingly emerged for S-NAT. Given that most listeners were US-based, we hypothesize that this shift was driven by an underlying ``model bias'': since most models were mainly trained on US English datasets, their synthesized outputs naturally align better with US listeners' perceptual expectations of naturalness.

Next we are interested in how listeners with different accents agree with each other. For each accent in the test samples, we calculated the system-level average scores for each listener group and calculated the Spearman's rank correlation ($\rho$). As shown in Table~\ref{tab:listener_agreement_by_accent_multi}, listeners generally exhibited moderate to strong agreement across most test accents ($\rho > 0.4$). While we initially hypothesized stronger disagreement when evaluating highly specific regional accents (e.g., Irish or Welsh), the current data did not provide sufficient evidence to support it.

\section{Conclusions}

This paper described the design and analysis of the CodecMOS-Accent dataset, which, to our knowledge, is the most extensive study to date regarding the generalization performance of NACs and NAC-based TTS models on accented speech, specifically in terms of the volume of samples, systems, listeners, and annotations. Our findings identify which systems fell short in subjective performance and demonstrate the extent to which current objective metrics align with human perception. Furthermore, this work provides insights into effective accent evaluation methodologies, while our bias analysis underscores the necessity of culturally diverse training data for achieving universally natural speech synthesis.

One of the most important future directions is to utilize this dataset to develop better SQA metrics. Recent advances in subjective speech quality predictors like UTMOS were built on large-scale listening test datasets like BVCC \cite{bvcc}, which, established in 2022, only comprised systems up to 2020. While we demonstrated that UTMOS scores correlate well with subjective naturalness, incorporating CodecMOS-Accent into future training sets will likely improve the robustness of SQA models. Finally, as current objective metrics for accent similarity rely largely on accent identification models, we believe that directly training models on human labels will lead to more accurate evaluation.

\ifcameraready

\section{Acknowledgments}
We would like to thank Henry Li Xinyuan from Johns Hopkins University for his fruitful comments on the dataset construction. This work was partly supported by JSPS KAKENHI Grant Number 25K00143.
\fi

\section{Generative AI Use Disclosure}

The authors used Gemini 3 Pro to (1) generate LaTex source codes for the colored tables (2) polish the English grammar and word usage. The authors take full responsibility for the contents in the manuscript.

\bibliographystyle{IEEEtran}
\bibliography{mybib}

\end{document}

%% file: tables/results.tex
\begin{table*}[t]
\footnotesize

\centering
\caption{System-level evaluation results comparing Ground Truth, resynthesis, and TTS models. The table reports subjective scores (S-NAT, S-SPK-SIM, S-ACC-SIM) as mean values $\pm$ $95\%$ confidence intervals, and objective metrics (O-WER, O-SPK-SIM, O-ACC-SIM, O-UTMOS) as mean values. Background color intensity indicates relative performance within each column, where darker blue denotes better performance (higher scores for all metrics except O-WER, for which lower is better). Systems are grouped by category and sorted by S-NAT score.}
\label{tab:results}

\begin{tabular}{lw{c}{1.8cm}w{c}{1.8cm}w{c}{1.8cm}w{c}{1.2cm}w{c}{1.2cm}w{c}{1.2cm}w{c}{1.2cm}}
\toprule
System & \textbf{S-NAT} & \textbf{S-SPK-SIM} & \textbf{S-ACC-SIM} & \textbf{O-WER} & \textbf{O-SPK-SIM} & \textbf{O-ACC-SIM} & \textbf{O-UTMOS} \\
\midrule
Ground Truth & \cellcolor[rgb]{0.381,0.686,0.890} 4.045 \(\pm\) 0.063 & \cellcolor[rgb]{0.204,0.596,0.859} 4.756 \(\pm\) 0.042 & \cellcolor[rgb]{0.204,0.596,0.859} 4.678 \(\pm\) 0.050 & \cellcolor[rgb]{0.418,0.705,0.897} 1.911 & \cellcolor[rgb]{0.204,0.596,0.859} 1.000 & \cellcolor[rgb]{0.204,0.596,0.859} 1.000 & \cellcolor[rgb]{0.456,0.724,0.904} 4.082 \\
\midrule
NanoCodec \cite{nanocodec} & \cellcolor[rgb]{0.368,0.679,0.888} 4.073 \(\pm\) 0.062 & \cellcolor[rgb]{0.286,0.638,0.873} 4.558 \(\pm\) 0.055 & \cellcolor[rgb]{0.299,0.644,0.876} 4.493 \(\pm\) 0.062 & \cellcolor[rgb]{0.273,0.631,0.871} 0.640 & \cellcolor[rgb]{0.351,0.671,0.885} 0.872 & \cellcolor[rgb]{0.342,0.666,0.883} 0.907 & \cellcolor[rgb]{0.424,0.708,0.898} 4.116 \\
FACodec \cite{naturalspeech3} & \cellcolor[rgb]{0.370,0.680,0.888} 4.070 \(\pm\) 0.061 & \cellcolor[rgb]{0.239,0.614,0.865} 4.671 \(\pm\) 0.048 & \cellcolor[rgb]{0.239,0.614,0.865} 4.609 \(\pm\) 0.053 & \cellcolor[rgb]{0.274,0.632,0.871} 0.650 & \cellcolor[rgb]{0.364,0.677,0.887} 0.861 & \cellcolor[rgb]{0.341,0.666,0.883} 0.908 & \cellcolor[rgb]{0.439,0.716,0.901} 4.100 \\
NeuCodec \cite{neucodec} & \cellcolor[rgb]{0.380,0.685,0.890} 4.048 \(\pm\) 0.064 & \cellcolor[rgb]{0.470,0.731,0.906} 4.116 \(\pm\) 0.077 & \cellcolor[rgb]{0.446,0.719,0.902} 4.207 \(\pm\) 0.074 & \cellcolor[rgb]{0.640,0.817,0.936} 3.847 & \cellcolor[rgb]{0.464,0.728,0.905} 0.774 & \cellcolor[rgb]{0.432,0.712,0.899} 0.847 & \cellcolor[rgb]{0.584,0.789,0.926} 3.945 \\
SNAC \cite{snac} & \cellcolor[rgb]{0.693,0.844,0.946} 3.367 \(\pm\) 0.080 & \cellcolor[rgb]{0.406,0.699,0.895} 4.268 \(\pm\) 0.069 & \cellcolor[rgb]{0.423,0.707,0.898} 4.252 \(\pm\) 0.071 & \cellcolor[rgb]{0.693,0.844,0.946} 4.318 & \cellcolor[rgb]{0.552,0.773,0.921} 0.697 & \cellcolor[rgb]{0.499,0.746,0.911} 0.802 & \cellcolor[rgb]{0.937,0.968,0.989} 3.567 \\
Mimi 4.4kbps \cite{moshi} & \cellcolor[rgb]{0.742,0.869,0.954} 3.261 \(\pm\) 0.074 & \cellcolor[rgb]{0.392,0.692,0.892} 4.302 \(\pm\) 0.066 & \cellcolor[rgb]{0.411,0.701,0.896} 4.276 \(\pm\) 0.069 & \cellcolor[rgb]{0.394,0.693,0.893} 1.700 & \cellcolor[rgb]{0.555,0.774,0.921} 0.695 & \cellcolor[rgb]{0.507,0.750,0.913} 0.796 & \cellcolor[rgb]{0.968,0.984,0.994} 3.534 \\
WavTokenizer 40hz \cite{wavtokenizer} & \cellcolor[rgb]{0.770,0.883,0.959} 3.200 \(\pm\) 0.077 & \cellcolor[rgb]{0.561,0.777,0.922} 3.896 \(\pm\) 0.078 & \cellcolor[rgb]{0.578,0.786,0.925} 3.951 \(\pm\) 0.078 & \cellcolor[rgb]{1.000,1.000,1.000} 11.735 & \cellcolor[rgb]{0.688,0.842,0.945} 0.579 & \cellcolor[rgb]{0.634,0.814,0.935} 0.711 & \cellcolor[rgb]{0.765,0.881,0.958} 3.751 \\
DAC 6kbps \cite{dac} & \cellcolor[rgb]{1.000,1.000,1.000} 1.418 \(\pm\) 0.055 & \cellcolor[rgb]{0.796,0.897,0.964} 3.329 \(\pm\) 0.086 & \cellcolor[rgb]{0.802,0.899,0.965} 3.516 \(\pm\) 0.089 & \cellcolor[rgb]{0.454,0.723,0.903} 2.220 & \cellcolor[rgb]{0.653,0.824,0.938} 0.610 & \cellcolor[rgb]{0.696,0.846,0.946} 0.669 & \cellcolor[rgb]{1.000,1.000,1.000} 1.928 \\
SpeechTokenizer 2 layers \cite{speechtokenizer} & \cellcolor[rgb]{1.000,1.000,1.000} 1.386 \(\pm\) 0.053 & \cellcolor[rgb]{1.000,1.000,1.000} 2.838 \(\pm\) 0.089 & \cellcolor[rgb]{1.000,1.000,1.000} 3.131 \(\pm\) 0.092 & \cellcolor[rgb]{0.878,0.938,0.978} 5.934 & \cellcolor[rgb]{1.000,1.000,1.000} 0.308 & \cellcolor[rgb]{0.930,0.965,0.988} 0.512 & \cellcolor[rgb]{1.000,1.000,1.000} 2.320 \\
Encodec 1.5kbps \cite{encodec} & \cellcolor[rgb]{1.000,1.000,1.000} 1.364 \(\pm\) 0.056 & \cellcolor[rgb]{0.773,0.885,0.960} 3.384 \(\pm\) 0.084 & \cellcolor[rgb]{0.796,0.897,0.964} 3.527 \(\pm\) 0.089 & \cellcolor[rgb]{0.671,0.833,0.942} 4.125 & \cellcolor[rgb]{0.768,0.882,0.959} 0.510 & \cellcolor[rgb]{0.746,0.871,0.955} 0.636 & \cellcolor[rgb]{1.000,1.000,1.000} 1.544 \\
\midrule
CosyVoice 2 \cite{cosyvoice2} & \cellcolor[rgb]{0.204,0.596,0.859} 4.430 \(\pm\) 0.052 & \cellcolor[rgb]{0.418,0.705,0.897} 4.241 \(\pm\) 0.071 & \cellcolor[rgb]{0.503,0.748,0.912} 4.097 \(\pm\) 0.082 & \cellcolor[rgb]{0.216,0.602,0.861} 0.143 & \cellcolor[rgb]{0.521,0.757,0.915} 0.724 & \cellcolor[rgb]{0.670,0.832,0.941} 0.687 & \cellcolor[rgb]{0.207,0.598,0.859} 4.349 \\
OpenAudio s1 mini \cite{openaudio} & \cellcolor[rgb]{0.268,0.629,0.870} 4.291 \(\pm\) 0.053 & \cellcolor[rgb]{0.452,0.722,0.903} 4.158 \(\pm\) 0.078 & \cellcolor[rgb]{0.488,0.740,0.909} 4.125 \(\pm\) 0.081 & \cellcolor[rgb]{0.211,0.599,0.860} 0.093 & \cellcolor[rgb]{0.645,0.820,0.937} 0.617 & \cellcolor[rgb]{0.704,0.850,0.947} 0.664 & \cellcolor[rgb]{0.211,0.600,0.860} 4.344 \\
Llasa-1B \cite{llasa} & \cellcolor[rgb]{0.291,0.640,0.874} 4.240 \(\pm\) 0.062 & \cellcolor[rgb]{0.462,0.727,0.905} 4.135 \(\pm\) 0.081 & \cellcolor[rgb]{0.548,0.771,0.920} 4.009 \(\pm\) 0.087 & \cellcolor[rgb]{0.211,0.599,0.860} 0.092 & \cellcolor[rgb]{0.695,0.845,0.946} 0.573 & \cellcolor[rgb]{0.781,0.889,0.961} 0.612 & \cellcolor[rgb]{0.204,0.596,0.859} 4.352 \\
Chatterbox \cite{chatterboxtts2025} & \cellcolor[rgb]{0.296,0.643,0.875} 4.230 \(\pm\) 0.058 & \cellcolor[rgb]{0.379,0.685,0.890} 4.333 \(\pm\) 0.071 & \cellcolor[rgb]{0.504,0.749,0.912} 4.094 \(\pm\) 0.085 & \cellcolor[rgb]{0.238,0.613,0.865} 0.328 & \cellcolor[rgb]{0.540,0.767,0.918} 0.708 & \cellcolor[rgb]{0.744,0.870,0.955} 0.637 & \cellcolor[rgb]{0.335,0.662,0.882} 4.212 \\
NeuTTS Air \cite{neuttsair} & \cellcolor[rgb]{0.374,0.682,0.889} 4.060 \(\pm\) 0.070 & \cellcolor[rgb]{0.820,0.909,0.968} 3.271 \(\pm\) 0.093 & \cellcolor[rgb]{0.782,0.889,0.961} 3.555 \(\pm\) 0.093 & \cellcolor[rgb]{0.287,0.638,0.874} 0.759 & \cellcolor[rgb]{0.786,0.891,0.962} 0.494 & \cellcolor[rgb]{0.793,0.895,0.963} 0.604 & \cellcolor[rgb]{0.250,0.619,0.867} 4.303 \\
IndexTTS2 \cite{indextts2} & \cellcolor[rgb]{0.412,0.702,0.896} 3.977 \(\pm\) 0.068 & \cellcolor[rgb]{0.298,0.644,0.875} 4.530 \(\pm\) 0.058 & \cellcolor[rgb]{0.425,0.708,0.898} 4.248 \(\pm\) 0.081 & \cellcolor[rgb]{0.204,0.596,0.859} 0.034 & \cellcolor[rgb]{0.500,0.746,0.911} 0.743 & \cellcolor[rgb]{0.652,0.823,0.938} 0.699 & \cellcolor[rgb]{0.490,0.741,0.910} 4.046 \\
TorToiSe \cite{tortoise} & \cellcolor[rgb]{0.430,0.711,0.899} 3.939 \(\pm\) 0.070 & \cellcolor[rgb]{0.644,0.819,0.937} 3.696 \(\pm\) 0.090 & \cellcolor[rgb]{0.830,0.914,0.970} 3.462 \(\pm\) 0.100 & \cellcolor[rgb]{0.214,0.601,0.861} 0.122 & \cellcolor[rgb]{0.838,0.918,0.971} 0.449 & \cellcolor[rgb]{1.000,1.000,1.000} 0.465 & \cellcolor[rgb]{0.459,0.726,0.904} 4.079 \\
VoiceStar \cite{voicestar} & \cellcolor[rgb]{0.474,0.733,0.907} 3.842 \(\pm\) 0.074 & \cellcolor[rgb]{0.353,0.672,0.885} 4.397 \(\pm\) 0.066 & \cellcolor[rgb]{0.402,0.697,0.894} 4.293 \(\pm\) 0.074 & \cellcolor[rgb]{1.000,1.000,1.000} 7.192 & \cellcolor[rgb]{0.668,0.831,0.941} 0.597 & \cellcolor[rgb]{0.741,0.869,0.954} 0.639 & \cellcolor[rgb]{0.323,0.656,0.880} 4.225 \\
XTTS \cite{xtts} & \cellcolor[rgb]{0.476,0.734,0.907} 3.838 \(\pm\) 0.067 & \cellcolor[rgb]{0.645,0.820,0.937} 3.694 \(\pm\) 0.090 & \cellcolor[rgb]{0.684,0.840,0.944} 3.745 \(\pm\) 0.090 & \cellcolor[rgb]{0.264,0.627,0.869} 0.559 & \cellcolor[rgb]{0.654,0.824,0.939} 0.609 & \cellcolor[rgb]{0.693,0.844,0.946} 0.671 & \cellcolor[rgb]{0.536,0.764,0.918} 3.997 \\
MaskGCT \cite{maskgct} & \cellcolor[rgb]{0.511,0.752,0.913} 3.763 \(\pm\) 0.070 & \cellcolor[rgb]{0.274,0.632,0.871} 4.586 \(\pm\) 0.056 & \cellcolor[rgb]{0.304,0.647,0.877} 4.483 \(\pm\) 0.061 & \cellcolor[rgb]{0.378,0.684,0.890} 1.556 & \cellcolor[rgb]{0.526,0.760,0.916} 0.720 & \cellcolor[rgb]{0.635,0.815,0.935} 0.710 & \cellcolor[rgb]{0.680,0.837,0.943} 3.843 \\
VevoTTS \cite{vevo} & \cellcolor[rgb]{0.558,0.776,0.922} 3.661 \(\pm\) 0.076 & \cellcolor[rgb]{0.317,0.653,0.879} 4.484 \(\pm\) 0.062 & \cellcolor[rgb]{0.374,0.682,0.889} 4.348 \(\pm\) 0.070 & \cellcolor[rgb]{0.816,0.906,0.967} 5.386 & \cellcolor[rgb]{0.565,0.779,0.923} 0.686 & \cellcolor[rgb]{0.682,0.838,0.944} 0.679 & \cellcolor[rgb]{0.651,0.823,0.938} 3.874 \\
FireRedTTS \cite{fireredtts} & \cellcolor[rgb]{0.583,0.788,0.926} 3.607 \(\pm\) 0.076 & \cellcolor[rgb]{0.757,0.877,0.957} 3.424 \(\pm\) 0.096 & \cellcolor[rgb]{0.850,0.924,0.973} 3.422 \(\pm\) 0.099 & \cellcolor[rgb]{0.334,0.662,0.882} 1.173 & \cellcolor[rgb]{0.824,0.911,0.969} 0.461 & \cellcolor[rgb]{0.921,0.960,0.986} 0.518 & \cellcolor[rgb]{0.526,0.760,0.916} 4.007 \\
MetaVoice \cite{metavoice} & \cellcolor[rgb]{0.828,0.913,0.970} 3.073 \(\pm\) 0.083 & \cellcolor[rgb]{0.493,0.743,0.910} 4.060 \(\pm\) 0.075 & \cellcolor[rgb]{0.529,0.761,0.916} 4.047 \(\pm\) 0.076 & \cellcolor[rgb]{0.588,0.791,0.927} 3.399 & \cellcolor[rgb]{0.709,0.852,0.948} 0.561 & \cellcolor[rgb]{0.826,0.912,0.969} 0.582 & \cellcolor[rgb]{0.957,0.978,0.992} 3.546 \\
VALL-E-X \cite{vall-e, valle-x} & \cellcolor[rgb]{0.828,0.913,0.970} 3.073 \(\pm\) 0.083 & \cellcolor[rgb]{0.577,0.785,0.925} 3.858 \(\pm\) 0.085 & \cellcolor[rgb]{0.663,0.829,0.940} 3.786 \(\pm\) 0.089 & \cellcolor[rgb]{1.000,1.000,1.000} 7.603 & \cellcolor[rgb]{0.773,0.885,0.960} 0.505 & \cellcolor[rgb]{0.885,0.942,0.980} 0.542 & \cellcolor[rgb]{0.861,0.929,0.975} 3.649 \\
Orpheus-TTS \cite{orpheus-tts} & \cellcolor[rgb]{0.964,0.982,0.994} 2.778 \(\pm\) 0.082 & \cellcolor[rgb]{0.871,0.935,0.977} 3.149 \(\pm\) 0.099 & \cellcolor[rgb]{0.882,0.940,0.979} 3.361 \(\pm\) 0.097 & \cellcolor[rgb]{0.306,0.648,0.877} 0.928 & \cellcolor[rgb]{0.924,0.961,0.987} 0.374 & \cellcolor[rgb]{0.918,0.958,0.985} 0.520 & \cellcolor[rgb]{0.861,0.929,0.975} 3.649 \\
\bottomrule
\end{tabular}
\end{table*}

%% file: tables/corr.tex
\begin{table}[t]
\centering
\caption{Pearson correlation coefficients of all metrics with respect to the subjective scores (S-NAT, S-SPK-SIM, S-ACC-SIM) in the utterance-level (top) and system-level (bottom). Color intensity denotes the magnitude.}
\label{tab:correlations}
\begin{tabular}{lccc}
\toprule
\textbf{Metric} & \textbf{S-NAT} & \textbf{S-SPK-SIM} & \textbf{S-ACC-SIM} \\
\midrule
\multicolumn{4}{l}{\textit{Utterance-level}} \\
S-NAT & \cellcolor[rgb]{0.204,0.596,0.859} 1.00 & & \\
S-SPK-SIM & \cellcolor[rgb]{0.754,0.875,0.956} 0.31 & \cellcolor[rgb]{0.204,0.596,0.859} 1.00 & \\
S-ACC-SIM & \cellcolor[rgb]{0.793,0.895,0.963} 0.26 & \cellcolor[rgb]{0.406,0.699,0.895} 0.75 & \cellcolor[rgb]{0.204,0.596,0.859} 1.00 \\
\midrule
O-WER & \cellcolor[rgb]{0.883,0.941,0.979} -0.15 & \cellcolor[rgb]{0.966,0.983,0.994} -0.04 & \cellcolor[rgb]{0.977,0.988,0.996} -0.03 \\
O-SPK-SIM & \cellcolor[rgb]{0.755,0.876,0.957} 0.31 & \cellcolor[rgb]{0.691,0.843,0.945} 0.39 & \cellcolor[rgb]{0.733,0.865,0.953} 0.34 \\
O-ACC-SIM & \cellcolor[rgb]{0.852,0.925,0.974} 0.19 & \cellcolor[rgb]{0.771,0.884,0.959} 0.29 & \cellcolor[rgb]{0.751,0.874,0.956} 0.31 \\
O-UTMOS & \cellcolor[rgb]{0.486,0.739,0.909} 0.65 & \cellcolor[rgb]{0.824,0.911,0.969} 0.22 & \cellcolor[rgb]{0.878,0.938,0.978} 0.15 \\
\midrule
\midrule
\multicolumn{4}{l}{\textit{System-level}} \\
S-NAT & \cellcolor[rgb]{0.204,0.596,0.859} 1.00 & & \\
S-SPK-SIM & \cellcolor[rgb]{0.474,0.733,0.907} 0.66 & \cellcolor[rgb]{0.204,0.596,0.859} 1.00 & \\
S-ACC-SIM & \cellcolor[rgb]{0.527,0.760,0.916} 0.59 & \cellcolor[rgb]{0.226,0.607,0.863} 0.97 & \cellcolor[rgb]{0.204,0.596,0.859} 1.00 \\
\midrule
O-WER & \cellcolor[rgb]{0.682,0.839,0.944} -0.40 & \cellcolor[rgb]{0.902,0.950,0.983} -0.12 & \cellcolor[rgb]{0.945,0.972,0.990} -0.07 \\
O-SPK-SIM & \cellcolor[rgb]{0.588,0.791,0.927} 0.52 & \cellcolor[rgb]{0.313,0.651,0.878} 0.86 & \cellcolor[rgb]{0.282,0.636,0.873} 0.90 \\
O-ACC-SIM & \cellcolor[rgb]{0.748,0.872,0.955} 0.32 & \cellcolor[rgb]{0.448,0.720,0.902} 0.69 & \cellcolor[rgb]{0.358,0.674,0.886} 0.81 \\
O-UTMOS & \cellcolor[rgb]{0.233,0.611,0.864} 0.96 & \cellcolor[rgb]{0.561,0.777,0.922} 0.55 & \cellcolor[rgb]{0.613,0.804,0.931} 0.49 \\
\bottomrule
\end{tabular}
\end{table}

%% file: tables/bias.tex
\begin{table}[t]
\centering
\caption{Mean subjective scores from listeners sharing the speaker's accent (Same Accent) and listeners with a different accent (Diff. Accent). Results are reported separately for ground truth only and the full dataset (``All''). $p$-values are derived from Welch's t-test.
}
\label{tab:bias}

\begin{tabular}{lrrr}
\toprule
\textbf{Metric} & \textbf{Same Accent} & \textbf{Diff. Accent} & \textbf{$p$-value} \\
\midrule
\multicolumn{4}{l}{\textit{Ground truth only}} \\
S-NAT & 4.145 & 4.026 & 0.160 \\
S-SPK-SIM & 4.855 & 4.737 & 0.011 \\
S-ACC-SIM & 4.831 & 4.650 & $<$ 0.001 \\
\midrule
\multicolumn{4}{l}{\textit{All}} \\
S-NAT & 3.616 & 3.489 & $<$ 0.001 \\
S-SPK-SIM & 4.124 & 3.997 & $<$ 0.001 \\
S-ACC-SIM & 4.236 & 3.955 & $<$ 0.001 \\
\bottomrule
\end{tabular}

\end{table}

%% file: tables/listener-group-corr-one-column.tex
\begin{table}[t]

\footnotesize

\centering
\caption{Spearman's $\rho$ broken down by test sample accent, with average and minimum values across all valid pairs of listener accent groups evaluating the same subset of systems. Only pairs with a statistically significant correlation ($p$-value $< 0.05$) are included.}
\label{tab:listener_agreement_by_accent}

\label{tab:listener_agreement_by_accent_multi}
\begin{tabular}{lcccccc}
\toprule
 & \multicolumn{2}{c}{\textbf{S-NAT}} & \multicolumn{2}{c}{\textbf{S-SPK-SIM}} & \multicolumn{2}{c}{\textbf{S-ACC-SIM}} \\
\cmidrule(lr){2-3} \cmidrule(lr){4-5} \cmidrule(lr){6-7}
\textbf{Speaker Accent} & \textbf{Avg} & \textbf{Min} & \textbf{Avg} & \textbf{Min} & \textbf{Avg} & \textbf{Min} \\
\midrule
South African & 0.90 & 0.88 & 0.77 & 0.65 & 0.78 & 0.64 \\
Australian & 0.86 & 0.84 & 0.71 & 0.59 & 0.68 & 0.61 \\
English & 0.86 & 0.84 & 0.72 & 0.59 & 0.75 & 0.71 \\
Northern Irish & 0.85 & 0.76 & 0.79 & 0.72 & 0.74 & 0.62 \\
American & 0.79 & 0.75 & 0.68 & 0.44 & 0.75 & 0.72 \\
Canadian & 0.79 & 0.71 & 0.70 & 0.44 & 0.75 & 0.64 \\
Scottish & 0.77 & 0.74 & 0.81 & 0.79 & 0.62 & 0.46 \\
New Zealand & 0.75 & 0.72 & 0.55 & 0.44 & 0.61 & 0.60 \\
Irish & 0.70 & 0.45 & 0.87 & 0.45 & 0.58 & 0.43 \\
Welsh & 0.52 & 0.48 & 0.59 & 0.52 & 0.59 & 0.51 \\
\bottomrule
\end{tabular}

\end{table}

%% file: mybib.bib
@article{arora2025on,
    title={{On The Landscape of Spoken Language Models: A Comprehensive Survey}},
    author={Siddhant Arora and Kai-Wei Chang and Chung-Ming Chien and Yifan Peng and Haibin Wu and Yossi Adi and Emmanuel Dupoux and Hung-yi Lee and Karen Livescu and Shinji Watanabe},
    journal={TMLR},
    issn={2835-8856},
    year={2025},
}

@ARTICLE{recent-advances-in-discrete-speech-tokens,
  author={Guo, Yiwei and Li, Zhihan and Wang, Hankun and Li, Bohan and Shao, Chongtian and Zhang, Hanglei and Du, Chenpeng and Chen, Xie and Liu, Shujie and Yu, Kai},
  journal={IEEE Trans. on Pattern Analysis and Machine Intelligence}, 
  title={{Recent Advances in Discrete Speech Tokens: A Review}}, 
  year={2025},
  pages={1-20},
}

@inproceedings{cui-etal-2025-recent,
    title = "Recent Advances in Speech Language Models: A Survey",
    author = "Cui, Wenqian  and
      Yu, Dianzhi  and
      Jiao, Xiaoqi  and
      Meng, Ziqiao  and
      Zhang, Guangyan  and
      Wang, Qichao  and
      Guo, Steven Y.  and
      King, Irwin",
    booktitle = "Proc. ACL",
    month = jul,
    year = "2025",
    address = "Vienna, Austria",
    pages = "13943--13970",
}

@ARTICLE{speech-ssl-review,
  author={Mohamed, Abdelrahman and Lee, Hung-yi and Borgholt, Lasse and Havtorn, Jakob D. and Edin, Joakim and Igel, Christian and Kirchhoff, Katrin and Li, Shang-Wen and Livescu, Karen and Maaløe, Lars and Sainath, Tara N. and Watanabe, Shinji},
  journal={IEEE Journal of Selected Topics in Signal Processing}, 
  title={{Self-Supervised Speech Representation Learning: A Review}}, 
  year={2022},
  volume={16},
  number={6},
  pages={1179-1210},
}

@article{moshi,
  title={Moshi: a speech-text foundation model for real-time dialogue},
  author={D{\'e}fossez, Alexandre and Mazar{\'e}, Laurent and Orsini, Manu and Royer, Am{\'e}lie and P{\'e}rez, Patrick and J{\'e}gou, Herv{\'e} and Grave, Edouard and Zeghidour, Neil},
  journal={arXiv preprint arXiv:2410.00037},
  year={2024}
}

@inproceedings{codec-superb,
    title = "Codec-{SUPERB}: An In-Depth Analysis of Sound Codec Models",
    author = "Wu, Haibin and Chung, Ho-Lam and Lin, Yi-Cheng and Wu, Yuan-Kuei and Chen, Xuanjun and Pai, Yu-Chi and Wang, Hsiu-Hsuan and Chang, Kai-Wei and Liu, Alexander and Lee, Hung-yi",
    booktitle = "Proc. Findings of the ACL",
    year = "2024",
    pages = "10330--10348",
}

@article{DASB,
  title={DASB -- Discrete Audio and Speech Benchmark},
  author={Mousavi, Pooneh and Della Libera, Luca and Duret, Jarod and Ploujnikov, Artem and Subakan, Cem and Ravanelli, Mirco},
  journal={arXiv preprint arXiv:2406.14294},
  year={2024}
}

@article{discrete-audio-tokens-more-than-a-survey,
    title={{Discrete Audio Tokens: More Than a Survey!}},
    author={Pooneh Mousavi and Gallil Maimon and Adel Moumen and Darius Petermann and Jiatong Shi and Haibin Wu and Haici Yang and Anastasia Kuznetsova and Artem Ploujnikov and Ricard Marxer and Bhuvana Ramabhadran and Benjamin Elizalde and Loren Lugosch and Jinyu Li and Cem Subakan and Phil Woodland and Minje Kim and Hung-yi Lee and Shinji Watanabe and Yossi Adi and Mirco Ravanelli},
    journal={TMLR},
    year={2025},
}

@ARTICLE{ac-tts-limited-data,
  author={Zhou, Xuehao and Zhang, Mingyang and Zhou, Yi and Wu, Zhizheng and Li, Haizhou},
  journal={IEEE/ACM TASLP}, 
  title={{Accented Text-to-Speech Synthesis With Limited Data}}, 
  year={2024},
  volume={32},
  pages={1699-1711},
}

@INPROCEEDINGS{accentbox,
  author={Zhong, Jinzuomu and Richmond, Korin and Su, Zhiba and Sun, Siqi},
  booktitle={Proc. ICASSP}, 
  title={{AccentBox: Towards High-Fidelity Zero-Shot Accent Generation}}, 
  year={2025},
  pages={1-5},
}

@inproceedings{pairwiase-evaluation-accent-sim,
  title     = {{Pairwise Evaluation of Accent Similarity in Speech Synthesis}},
  author    = {Jinzuomu Zhong and Suyuan Liu and Dan Wells and Korin Richmond},
  year      = {2025},
  booktitle = {{Proc. Interspeech}},
  pages     = {2290--2294},
}

@inproceedings{scalable-controllable-accented-tts,
  title={Scalable Controllable Accented TTS},
  author={Xinyuan, Henry Li and Cai, Zexin and Garg, Ashi and Duh, Kevin and Garc{\'\i}a-Perera, Leibny Paola and Khudanpur, Sanjeev and Andrews, Nicholas and Wiesner, Matthew},
  booktitle={Proc. ASRU},
  year={2025}
}

@inproceedings{snac,
    title={{SNAC}: Multi-Scale Neural Audio Codec},
    author={Hubert Siuzdak and Florian Gr{\"o}tschla and Luca A Lanzend{\"o}rfer},
    booktitle={Audio Imagination: NeurIPS 2024 Workshop AI-Driven Speech, Music, and Sound Generation},
    year={2024},
}

@inproceedings{dac,
     author = {Kumar, Rithesh and Seetharaman, Prem and Luebs, Alejandro and Kumar, Ishaan and Kumar, Kundan},
     booktitle = {Proc. NeurIPS},
     pages = {27980--27993},
     title = {High-Fidelity Audio Compression with Improved RVQGAN},
     volume = {36},
     year = {2023}
}

@article{encodec,
    title={High Fidelity Neural Audio Compression},
    author={Alexandre D{\'e}fossez and Jade Copet and Gabriel Synnaeve and Yossi Adi},
    journal={TMLR},
    year={2023},
}

@InProceedings{naturalspeech3,
  title = 	 {{N}atural{S}peech 3: Zero-Shot Speech Synthesis with Factorized Codec and Diffusion Models},
  author =       {Ju, Zeqian and Wang, Yuancheng and Shen, Kai and Tan, Xu and Xin, Detai and Yang, Dongchao and Liu, Eric and Leng, Yichong and Song, Kaitao and Tang, Siliang and Wu, Zhizheng and Qin, Tao and Li, Xiangyang and Ye, Wei and Zhang, Shikun and Bian, Jiang and He, Lei and Li, Jinyu and Zhao, Sheng},
  booktitle = 	 {Proc. ICML},
  pages = 	 {22605--22623},
  year = 	 {2024},
  volume = 	 {235},
  month = 	 {21--27 Jul},
}

@inproceedings{nanocodec,
  title     = {{NanoCodec: Towards High-Quality Ultra Fast Speech LLM Inference}},
  author    = {Edresson Casanova and Paarth Neekhara and Ryan Langman and Shehzeen Hussain and Subhankar Ghosh and Xuesong Yang and Ante Jukic and Jason Li and Boris Ginsburg},
  year      = {2025},
  booktitle = {{Proc. Interspeech}},
  pages     = {5028--5032},
}

@article{neucodec,
  title={Finite Scalar Quantization Enables Redundant and Transmission-Robust Neural Audio Compression at Low Bit-rates},
  author={Julian, Harry and Beeson, Rachel and Konathala, Lohith and Ulin, Johanna and Gao, Jiameng},
  journal={arXiv preprint arXiv:2509.09550},
  year={2025},
}

@inproceedings{speechtokenizer,
    title={{SpeechTokenizer: Unified Speech Tokenizer for Speech Language Models}},
    author={Xin Zhang and Dong Zhang and Shimin Li and Yaqian Zhou and Xipeng Qiu},
    booktitle={Proc. ICLR},
    year={2024},
}

@inproceedings{wavtokenizer,
    title={{WavTokenizer: an Efficient Acoustic Discrete Codec Tokenizer for Audio Language Modeling}},
    author={Shengpeng Ji and Ziyue Jiang and Wen Wang and Yifu Chen and Minghui Fang and Jialong Zuo and Qian Yang and Xize Cheng and Zehan Wang and Ruiqi Li and Ziang Zhang and Xiaoda Yang and Rongjie Huang and Yidi Jiang and Qian Chen and Siqi Zheng and Zhou Zhao},
    booktitle={Proc. ICLR},
    year={2025},
}

@ARTICLE{soundstream,
  author={Zeghidour, Neil and Luebs, Alejandro and Omran, Ahmed and Skoglund, Jan and Tagliasacchi, Marco},
  journal={IEEE/ACM TASLP}, 
  title={{SoundStream: An End-to-End Neural Audio Codec}}, 
  year={2022},
  volume={30},
  pages={495-507},
}

@inproceedings{neural-voice-cloning,
    author = {Arik, Sercan and Chen, Jitong and Peng, Kainan and Ping, Wei and Zhou, Yanqi},
    booktitle = {Proc. NIPS},
    title = {{Neural Voice Cloning with a Few Samples}},
    volume = {31},
    year = {2018}
}

@ARTICLE{vall-e,
  author={Chen, Sanyuan and Wang, Chengyi and Wu, Yu and Zhang, Ziqiang and Zhou, Long and Liu, Shujie and Chen, Zhuo and Liu, Yanqing and Wang, Huaming and Li, Jinyu and He, Lei and Zhao, Sheng and Wei, Furu},
  journal={IEEE TASLP}, 
  title={{Neural Codec Language Models are Zero-Shot Text to Speech Synthesizers}}, 
  year={2025},
  volume={33},
  pages={705-718},
}

@article{cosyvoice3,
  title={{Cosyvoice 3: Towards in-the-wild speech generation via scaling-up and post-training}},
  author={Du, Zhihao and Gao, Changfeng and Wang, Yuxuan and Yu, Fan and Zhao, Tianyu and Wang, Hao and Lv, Xiang and Wang, Hui and Ni, Chongjia and Shi, Xian and others},
  journal={arXiv preprint arXiv:2505.17589},
  year={2025}
}

@misc{chatterboxtts2025,
  author       = {{Resemble AI}},
  title        = {{Chatterbox-TTS}},
  year         = {2025},
  howpublished = {\url{https://github.com/resemble-ai/chatterbox}},
  note         = {GitHub repository}
}

@article{cosyvoice2,
  title={{Cosyvoice 2: Scalable streaming speech synthesis with large language models}},
  author={Du, Zhihao and Wang, Yuxuan and Chen, Qian and Shi, Xian and Lv, Xiang and Zhao, Tianyu and Gao, Zhifu and Yang, Yexin and Gao, Changfeng and Wang, Hui and others},
  journal={arXiv preprint arXiv:2412.10117},
  year={2024}
}

@article{fireredtts,
  title={{FireRedTTS: A foundation text-to-speech framework for industry-level generative speech applications}},
  author={Guo, Hao-Han and Hu, Yao and Liu, Kun and Shen, Fei-Yu and Tang, Xu and Wu, Yi-Chen and Xie, Feng-Long and Xie, Kun and Xu, Kai-Tuo},
  journal={arXiv preprint arXiv:2409.03283},
  year={2024}
}

@article{indextts2,
  title={{IndexTTS2: A Breakthrough in Emotionally Expressive and Duration-Controlled Auto-Regressive Zero-Shot Text-to-Speech}},
  author={Zhou, Siyi and Zhou, Yiquan and He, Yi and Zhou, Xun and Wang, Jinchao and Deng, Wei and Shu, Jingchen},
  journal={arXiv preprint arXiv:2506.21619},
  year={2025}
}

@article{llasa,
  title={{Llasa: Scaling Train-Time and Inference-Time Compute for Llama-based Speech Synthesis}},
  author={Ye, Zhen and Zhu, Xinfa and Chan, Chi-Min and Wang, Xinsheng and Tan, Xu and Lei, Jiahe and Peng, Yi and Liu, Haohe and Jin, Yizhu and Dai, Zheqi and others},
  journal={arXiv preprint arXiv:2502.04128},
  year={2025}
}

@inproceedings{maskgct,
    title={Mask{GCT}: Zero-Shot Text-to-Speech with Masked Generative Codec Transformer},
    author={Yuancheng Wang and Haoyue Zhan and Liwei Liu and Ruihong Zeng and Haotian Guo and Jiachen Zheng and Qiang Zhang and Xueyao Zhang and Shunsi Zhang and Zhizheng Wu},
    booktitle={Proc. ICLR},
    year={2025},
}

@misc{metavoice,
  author       = {{MetaVoice}},
  title        = {{MetaVoice}},
  year         = {2024},
  howpublished = {\url{https://github.com/metavoiceio/metavoice-src}},
  note         = {GitHub repository}
}

@misc{neuttsair,
  author       = {{Neuphonic}},
  title        = {{NeuTTS Air}},
  year         = {2025},
  howpublished = {\url{https://github.com/neuphonic/neutts/tree/main}},
  note         = {GitHub repository}
}

@article{openaudio,
    title={{Fish-Speech: Leveraging Large Language Models for Advanced Multilingual Text-to-Speech Synthesis}},
    author={Shijia Liao and Yuxuan Wang and Tianyu Li and Yifan Cheng and Ruoyi Zhang and Rongzhi Zhou and Yijin Xing},
    year={2024},
    journal={arXiv preprint arXiv:2411.01156},
}

@misc{orpheus-tts,
  author       = {{Canopy AI}},
  title        = {{Orpheus-TTS}},
  year         = {2025},
  howpublished = {\url{https://github.com/canopyai/Orpheus-TTS}},
  note         = {GitHub repository}
}

@article{tortoise,
  title={Better speech synthesis through scaling},
  author={Betker, James},
  journal={arXiv preprint arXiv:2305.07243},
  year={2023}
}

@misc{valle-x,
  author       = {{Plachtaa}},
  title        = {{VALL-E-X}},
  year         = {2023},
  howpublished = {\url{https://github.com/Plachtaa/VALL-E-X}},
  note         = {GitHub repository}
}

@inproceedings{vevo,
  author       = {Xueyao Zhang and Xiaohui Zhang and Kainan Peng and Zhenyu Tang and Vimal Manohar and Yingru Liu and Jeff Hwang and Dangna Li and Yuhao Wang and Julian Chan and Yuan Huang and Zhizheng Wu and Mingbo Ma},
  title        = {{Vevo: Controllable Zero-Shot Voice Imitation with Self-Supervised Disentanglement}},
  booktitle    = {{Proc. ICLR}},
  year         = {2025}
}

@article{voicestar,
  title={{VoiceStar: Robust Zero-Shot Autoregressive TTS with Duration Control and Extrapolation}},
  author={Peng, Puyuan and Li, Shang-Wen and Mohamed, Abdelrahman and Harwath, David},
  journal={arXiv preprint arXiv:2505.19462},
  year={2025}
}

@inproceedings{xtts,
  title     = {{XTTS: a Massively Multilingual Zero-Shot Text-to-Speech Model}},
  author    = {Edresson Casanova and Kelly Davis and Eren Gölge and Görkem Göknar and Iulian Gulea and Logan Hart and Aya Aljafari and Joshua Meyer and Reuben Morais and Samuel Olayemi and Julian Weber},
  year      = {2024},
  booktitle = {{Proc. Interspeech}},
  pages     = {4978--4982},
}

@inproceedings{naturalspeech2,
    title={{NaturalSpeech 2: Latent Diffusion Models are Natural and Zero-Shot Speech and Singing Synthesizers}},
    author={Kai Shen and Zeqian Ju and Xu Tan and Eric Liu and Yichong Leng and Lei He and Tao Qin and sheng zhao and Jiang Bian},
    booktitle={Proc. ICLR},
    year={2024},
}

@INPROCEEDINGS{factorized-tacotron,
  author={Hsu, Wei-Ning and Zhang, Yu and Weiss, Ron J. and Chung, Yu-An and Wang, Yuxuan and Wu, Yonghui and Glass, James},
  booktitle={Proc. ICASSP}, 
  title={{Disentangling Correlated Speaker and Noise for Speech Synthesis via Data Augmentation and Adversarial Factorization}}, 
  year={2019},
  pages={5901-5905},
}

@misc{vctk,
  title="{CSTR VCTK Corpus: English Multi-speaker Corpus for CSTR Voice Cloning Toolkit}",
  author={C. Veaux and J. Yamagishi and Kirsten MacDonald},
  year={2017}
}

@inproceedings{bvcc,
  author={E. Cooper and J. Yamagishi},
  title={How do Voices from Past Speech Synthesis Challenges Compare Today?},
  year=2021,
  booktitle={Proc. 11th ISCA Speech Synthesis Workshop (SSW 11)},
  pages={183--188},
}

@InProceedings{whisper,
  title = 	 {{Robust Speech Recognition via Large-Scale Weak Supervision}},
  author =       {Radford, Alec and Kim, Jong Wook and Xu, Tao and Brockman, Greg and Mcleavey, Christine and Sutskever, Ilya},
  booktitle = 	 {Proc. ICML},
  pages = 	 {28492--28518},
  year = 	 {2023},
  volume = 	 {202},
  month = 	 {23--29 Jul},
}

@inproceedings{ecapa-tdnn,
  author    = {Brecht Desplanques and
               Jenthe Thienpondt and
               Kris Demuynck},
  editor    = {Helen Meng and
               Bo Xu and
               Thomas Fang Zheng},
  title     = {{ECAPA-TDNN:} Emphasized Channel Attention, Propagation and Aggregation
               in {TDNN} Based Speaker Verification},
  booktitle = {Proc. Interspeech},
  pages     = {3830--3834},
  year      = {2020},
}

@article{commonaccent,
  title={{CommonAccent: Exploring Large Acoustic Pretrained Models for Accent Classification Based on Common Voice}},
  author={Zuluaga-Gomez, Juan and Ahmed, Sara and Visockas, Danielius and Subakan, Cem},
  pages = {5291--5295},
  journal={Proc. Interspeech},
  year={2023}
}

@inproceedings{utmos,
  author={T. Saeki and D. Xin and W. Nakata and T. Koriyama and S. Takamichi and H. Saruwatari},
  title={{UTMOS: UTokyo-SaruLab System for VoiceMOS Challenge 2022}},
  year=2022,
  booktitle={Proc. Interspeech},
  pages={4521--4525},
}

@inproceedings{voicemos2022,
  author={W.-C. Huang and E. Cooper and Y. Tsao and H.-M. Wang and T. Toda and J. Yamagishi},
  title={{The VoiceMOS Challenge 2022}},
  year=2022,
  booktitle={Proc. Interspeech},
  pages={4536--4540},
}

@inproceedings{obj_vcc2020,
  title     = {{Predictions of Subjective Ratings and Spoofing Assessments of Voice Conversion Challenge 2020 Submissions}},
  author    = {Rohan Kumar Das and Tomi Kinnunen and Wen-Chin Huang and Zhen-Hua Ling and Junichi Yamagishi and Zhao Yi and Xiaohai Tian and Tomoki Toda},
  year      = {2020},
  booktitle = {{Joint Workshop for the Blizzard Challenge and Voice Conversion Challenge 2020}},
  pages     = {99--120},
  doi       = {10.21437/VCCBC.2020-15},
}
